\documentclass[conference]{IEEEtran}

\usepackage{graphicx}
\usepackage[space]{grffile}
\usepackage{latexsym}
\usepackage{textcomp}
\usepackage{longtable}
\usepackage{multirow,booktabs}
\usepackage{amsfonts,amsmath,amssymb}
\usepackage{url}
\usepackage{hyperref}
\hypersetup{colorlinks=false,pdfborder={0 0 0}}
\newif\iflatexml\latexmlfalse
\usepackage[utf8]{inputenc}
\usepackage[english]{babel}
\usepackage{tabu}
\newcommand\citet{\cite}

\begin{document}


\title{SHA-1 and the Strict Avalanche Criterion}


%
%

\author{\IEEEauthorblockN{Yusuf Motara}
\IEEEauthorblockA{Rhodes University\\
Grahamstown 6140\\
SOUTH AFRICA\\
Email: y.motara@ru.ac.za}
\and
\IEEEauthorblockN{Barry Irwin}
\IEEEauthorblockA{Rhodes University\\
Grahamstown 6140\\
SOUTH AFRICA\\
Email: b.irwin@ru.ac.za}}


%
%

\maketitle

\begin{abstract}
The Strict Avalanche Criterion (SAC) is a measure of both confusion and diffusion, which are key properties of a cryptographic hash function.  This work provides a working definition of the SAC, describes an experimental methodology that can be used to statistically evaluate whether a cryptographic hash meets the SAC, and uses this to investigate the degree to which compression function of the SHA-1 hash meets the SAC.  The results ($P < 0.01$) are heartening: SHA-1 closely tracks the SAC after the first 24 rounds, and demonstrates excellent properties of confusion and diffusion throughout.

\end{abstract}

\section{Introduction}

Many computer scientists know little about the inner workings of cryptographic hashes, though they may know something about their properties.  One of these properties is the ``avalanche effect'', by analogy with the idea of a small stone causing a large avalanche of changes.  The ``avalanche effect'' explains how a small change in the input data can result in a large change in the output hash.  However, many questions around the effect are unanswered.  For example, how large is the effect?  After how many ``rounds'' of a compression function can it be seen?  Do all inputs result in such an effect?  Little experimental work has been done to answer these questions for any hash function, and this paper contributes experimental results that help in this regard.

A boolean $n$-bit hash function $H$ is the transform $\mathbb{Z}_{2}^m \to \mathbb{Z}_{2}^n$.  A cryptographic hash function attempts to obscure the relationship between the input and output of $F$, and the degree to which this is accomplished is directly related to the (second-)preimage resistance of the hash function.  This implies that two similar inputs should have very different outputs.

The Strict Avalanche Criterion (SAC)~(\cite{Webster_1986,Forri__1990}) formalizes this notion by measuring the amount of change introduced in the output by a small change in the input.  It builds on the definition of \emph{completeness}, which means that each bit of the output depends on all the bits of the input, in a way that is cryptographically relevant.  Using the definition of $H$ as above, an output $H(x) = y$ is obtained for an input $x$.  The initial bit of $x$ is now flipped, giving $H(x_0) = y_0$.  This process is repeated for $x_{1..n}$, resulting in $y_{1..n}$.  The SAC is met when the Hamming distance between $y$ and $y_{0..n}$ is, on average, $\frac{n}{2}$.


There are three contributions that this paper makes to the existing body of research:

\begin{enumerate}
\item A definition of what the SAC is;
\item Experimental SAC results for a particular cryptographic hash (SHA-1);
\item An exploration of intermediate results
\end{enumerate}

Section 2 of this paper examines related work and argues that the SAC as proposed by Webster \& Tavares~\cite{Webster_1986} has been misunderstood in much of the contemporaneous critical literature.  Section 3 introduces salient points of a well-known cryptographic hash (SHA-1) which is assumed to exhibit the SAC, and describes an experimental design to test its SAC-compliance.  Section 4 presents experimental results, and some discussion follows.

\section{Related work}

The original definition~\cite{Webster_1986} of the SAC is:

\begin{quotation}
Consider $X$ and $X_i$, two n-bit, binary plaintext vectors, such that $X$ and $X_i$ differ only in bit $i$, $1 < i < n$. Let

$$V_i = Y \oplus Y_i$$

where $Y = f(X)$ , $Y_i = f(X_i)$ and f is the cryptographic transformation, under consideration.  If $f$ is to meet the strict avalanche criterion, the probability that each bit in $V_i$ is equal to 1 should be one half over the set of all possible plaintext vectors $X$ and $X_i$. This should be true for all values of $i$. 
\end{quotation}

Forr\'e~\cite{Forri__1990} expresses this as:

\begin{quotation}
Let $\underline{x}$ and $\underline{x}_i$ denote two $n$-bit vectors, such that $\underline{x}$ and $\underline{x}_i$ differ only in bit $i$, $1 \leq i \leq n$.  $Z_{2}^{n}$ denotes the $n$-dimensional vector space over {0,1}.  The function $f(\underline{x}) = z, z \in \{0,1\}$ fulfills the SAC if and only if

$$\sum_{\underline{x} \in Z_2^n} f(\underline{x}) \oplus f(\underline{x}_i) = 2^{n-1}, \text{ for all }i\text{ with }1 \leq i \leq n.$$
\end{quotation}

Similarly, Lloyd~\cite{Lloyd_1990} understands the SAC as:

\begin{quotation}
Let $f : Z_2^n \mapsto Z_2^m$ be a cryptographic transformation.  Then $f$ satisfies the strict avalanche criterion if and only if

$$\sum_{\underline{x} \in Z_2^n}f(\underline{x}) \oplus f(\underline{x} \oplus \underline{c}_i) = (2^{n-1},...,2^{n-1}) \text{ for all }i\text{, }1 \leq i \leq n.$$

where $\oplus$ denotes bitwise exclusive or and $c_i$ is a vector of length $n$ with a 1 in the $i$th position and 0 elsewhere.
\end{quotation}

Other works (\cite{preneel1993analysis,Lloyd_1993,Babbage_1990,Kim_1991}) follow in the same vein.  However, these definitions calculate the sum over \emph{all} possible inputs as leading to the fulfillment of the SAC, which is contrary to the original definition.  The original definition separates a \emph{baseline} value from the \emph{avalanche vectors}, and states that the SAC holds true when ``the probability that each bit [in the avalanche vectors] is equal to 1 should be one half over the set of all possible plaintext vectors''~\cite{Webster_1986}.  Therefore, a better test of whether $f : \mathbb{Z}_2^n \mapsto \mathbb{Z}_2$ fulfills the SAC would use a universal quantifier,

\begin{gather*}
\forall \underline{x} \in \mathbb{Z}_2^n, P(f(\underline{x}) = f(\underline{x}_i)) = 0.5 \\ \text{for all }\underline{x}_i\text{ which differ from }\underline{x}\text{ in bit }i, 1 \leq i \leq n
\end{gather*}

A simple example clarifies the difference.  Babbage~\cite{Babbage_1990} uses Lloyd's~\cite{Lloyd_1990} definition of the SAC and defines a SAC-compliant function:

\begin{quotation}
Define $f : Z_2^n \mapsto Z_2$ by

\begin{equation*}
\begin{cases}
f(x_1,...,x_n) = 0 & \text{if }x_1 = 0 \\
f(x_1,...,x_n) = x_2 \oplus ... \oplus x_n & \text{if }x_1 = 1
\end{cases}
\end{equation*}
\end{quotation}

The simplest function of this nature is $f(\underline{x}) = x_0 \land x_1$.  Then, taking $g(\underline{x}) = f(\underline{x}) \oplus f(\underline{x} \oplus 01)$ and $h(\underline{x}) = f(\underline{x}) \oplus f(\underline{x} \oplus 10)$,

\begin{center}
\begin{tabu}{c|c|c|c|c}
$\underline{x}$ & $f(\underline{x})$ & $g(\underline{x})$ & $h(\underline{x})$ & $P(f(\underline{x}) = f(\underline{x}_i))$ \\ \hline
00 & 0 & 0 & 0 & 1.0 \\
01 & 0 & 0 & 1 & 0.5 \\
10 & 0 & 1 & 0 & 0.5 \\
11 & 1 & 1 & 1 & 1.0 \\
\hline
\multicolumn{2}{r}{Sum:} & \multicolumn{1}{c}{2} & \multicolumn{2}{l}{2} \\
\end{tabu}
\end{center}

Note that the sum of each of the third and fourth columns is $2^{n-1}$, as predicted, and that this function fulfills the summed definition of the SAC.  However, the first and last rows do not fulfill the original definition of the SAC at all: the probability of change, given the baseline values 00 and 11, is 0.0 in each case.  It is therefore more reasonable to regard the \emph{row} probability as important.  This understanding is also in accordance with the original text that defined the term.  Under this definition, $x_0 \land x_1$ is not SAC-compliant.

It is worth noting that the original definition, as per Webster \& Tavares~\cite{Webster_1986}, is slightly ambiguous.  They state that ``the probability that each bit in $V_i$ is equal to 1 should \emph{be one half} over the set of all possible plaintext vectors $X$ and $X_i$''; however, they also state that ``to satisfy the strict avalanche criterion, every element must have a value \emph{close to one half}'' (emphasis mine).  Under Lloyd's interpretation, the SAC is only satisfied when an element changes with a probability of precisely 0.5.  This is an unnecessarily binary criterion, as it seems to be more useful (and more in line with the original definition) to understand how far a particular sample \emph{diverges} from the SAC.  Therefore, this paper regards the SAC as a continuum but takes Lloyd's formulation as the definition of what it means to ``meet'' the SAC.

Preneel~\cite{preneel1993analysis} suggests a generalisation of the SAC called the \emph{propagation criterion} (PC), defined as

\begin{quotation}
Let $f$ be a Boolean function of $n$ variables.  Then $f$ satisfies the \textbf{propagation criterion of degree} $k$, $PC(k)$, ($1 \leq k \leq n$), if $\hat{f}(\underline{x})$ changes with a probability of 1/2 whenever $i$ ($1 \leq i \leq k$) bits of $\underline{x}$ are complemented.
\end{quotation}

It can be seen that the SAC is equivalent to $PC(1)$.  The same work defines an \emph{extended propagation criterion} which regards the SAC as a continuum.  Much of the subsequent work (\cite{Seberry_1994,Zhang_1996,Carlet_1998,Sung_1999,Canteaut_2000,Gouget_2004}) in this area has more closely examined the relationship between PC and nonlinearity characteristics.  Many of these extend the PC in interesting ways and examine ways of constructing functions which satisfy $PC(n)$, but experimental research that targets existing algorithms is scarce.

Although there are proven theoretical ways to construct a function which satisfies the SAC~\cite{Kim_1991}, there is no way (apart from exhaustive testing) to verify that an existing function satisfies the SAC.  By contrast, useful cryptographic properties such as non-degeneracy~\cite{dubuc2001characterization} or bentness~\cite{Rothaus_1976} are verifiable without having to resort to exhaustive testing.  However, the SAC metric is no worse in this regard than the correlation immunity~\citet{siegenthaler1984correlation} and balance~\cite{staffelbach1991cryptographic} metrics which also require exhaustive testing.

\section{Experimental design}

The SHA-1 hash~\cite{fips1995180} is a well-known cryptographic hash function which generates a 160-bit hash value. It is the successor to the equally well-known MD5 cryptographic hash function which generated a 128-bit hash value.  SHA-1 was designed by the National Security Agency of the United States of America and published in 1995 as National Institute of Standards and Technology (NIST) Federal Information Processing Standard 180-1.

\subsection{Hash details}

The SHA-1 hash is constructed using the Merkle-D\r{a}mgard paradigm~(\cite{merkle1979secrecy,gauravaram2005some}), which means that it consists of padding, chunking, and compression stages.  These stages are necessary for the hash algorithm to be able to handle inputs which are greater than 447 bits in length; however, they are unnecessary to consider in an examination of the compression function itself, since the strength of the Merkle-D\r{a}mgard paradigm is predicated on the characteristics of the compression function.  This paper examines only the compression function itself, and does not concern itself with padding, chunking, or Davies-Meyer strengthening~\cite{winternitz1984secure}.

The SHA-1 compression function makes use of addition, rotation, and logical functions (AND, OR, NOT, XOR), applied over the course of 80 rounds, to convert the 16 input words into a 5-word (160-bit) output.  Each round affects the calculation of subsequent rounds, and the hashing process can therefore not be parallelized to any significant degree.  A full description of the inner workings of SHA-1 is provided in FIPS 180-1~\cite{fips1995180}.  For the purposes of this work, it is sufficient to understand that each round generates a value that is used in subsequent rounds, and that there are 80 rounds in total.

\subsection{Statistical approach}

It is computationally infeasible to exhaustively test the degree to which SHA-1 meets the SAC since the input space ($2^{672}$) is too large.  However, it is possible to use a sampling approach instead, where representative samples are drawn from a \emph{population} and inferences are made based on an analysis of those samples.  This approach relies on each input being statistically independent of other inputs.  Generating such input can be an extraordinarily difficult task~\cite{Chaitin_2001}; however, random.org\footnote{\url{https://www.random.org}} provides data which meets this requirement~\cite{kenny2005random,foley2001analysis}.  A source which may be more random, but which has undergone far less scrutiny, is HotBits\footnote{\url{https://www.fourmilab.ch/hotbits/}}.  Data for these experiments has therefore been obtained from random.org.

The inputs which make up the population should represent real-world usage, and the form of the input is therefore of concern.  The inputs to the SHA-1 compression function are twofold: 16 32-bit words of input data and an initialization vector of 5 32-bit words, for a total of 21 32-bit words (or 672 bytes).  The initial initialization vector is defined by the FIPS 180-1 specification, and the input data is padded and terminated such that the last two words processed by the algorithm encode the length of the input data.  Subsequent initialization vectors are generated from the output of the previous application of the compression function.  For any input which is larger than 1024 bytes, there is therefore at least one iteration of the compression function for which all 672 bytes are effectively "random" --- if it is assumed that a previous iteration of the compression function can possibly result in the applicable initialization vector.  To make this assumption, is sufficient to assert that there are no values which \emph{cannot} be generated as intermediate intitialization vectors (given a pool of $\leq\!\!2^{64}$ different bitstreams).  Therefore, we can take independent 672-byte inputs as our population of concern.

The hypothesis to be tested is that SHA-1 meets the SAC.  The desired margin of error is 1\%, at a 99\% confidence level.  The required sample size  is therefore determined by

\begin{align*}
n = \left(\frac{\textrm{erf}^{-1}(0.99)}{0.01\sqrt{2}}\right)^2 = 16587
\\
\textrm{where erf}^{-1}\textrm{ is the inverse error function}
\end{align*}

Given the $2^{672}$ input space, this seems to be a very small number; however, ``it is the absolute size of the sample which determines accuracy, not the size relative to the population''~\cite{freedman2007statistics}.  Data collected during the experiment also indicates the degree to which SHA-1 does not meet the SAC, and the round at which the SAC comes into effect.

Each of the 16587 inputs is passed through a custom implementation of the SHA-1 compression function.  This implementation has not been validated by NIST's Cryptographic Algorithm Validation Program\footnote{\url{http://csrc.nist.gov/groups/STM/cavp/\#03}}, but nevertheless passes all of the byte-oriented test vectors provided by NIST; in addition, source code for the compression function is available on request.  When presented with a 672-byte input, the compression function outputs a list of 80 vectors, one for each round of the compression function.  \emph{Baseline} and \emph{avalanche} vectors are generated for each input, and per-round compliance with the SAC is determined by these.

The primary question that this work seeks to answer is: to what degree do each of the output bits meet the SAC?  To determine this, the per-input SAC value for each bit must be calculated, as described above.  The geometric mean of the SAC values is representative of the central tendency.  From the data that is generated to answer the primary question, two other questions may be fruitfully answered:

\begin{itemize}
\item \textbf{What is the distribution of SAC values per input?}  The geometric mean provides a way to understand the degree to which an output bit meets the SAC, on average over a range of inputs.  The distribution of SAC values quantifies how likely any particular input is to meet the SAC.
\item \textbf{How quickly do the bits of the SHA-1 hash meet (or not meet) the SAC?}
\end{itemize}

For repeatability, it is disclosed that the data used to create inputs is the first $16587 \times 672 = 11,146,464$ bits generated by random.org from the 2\textsuperscript{nd} to the 12\textsuperscript{th} of January 2015.  This data is available from \url{https://www.random.org/files/}.

\section{Results}

\begin{figure}[h]
	\centering
	\includegraphics[width=0.5\textwidth]{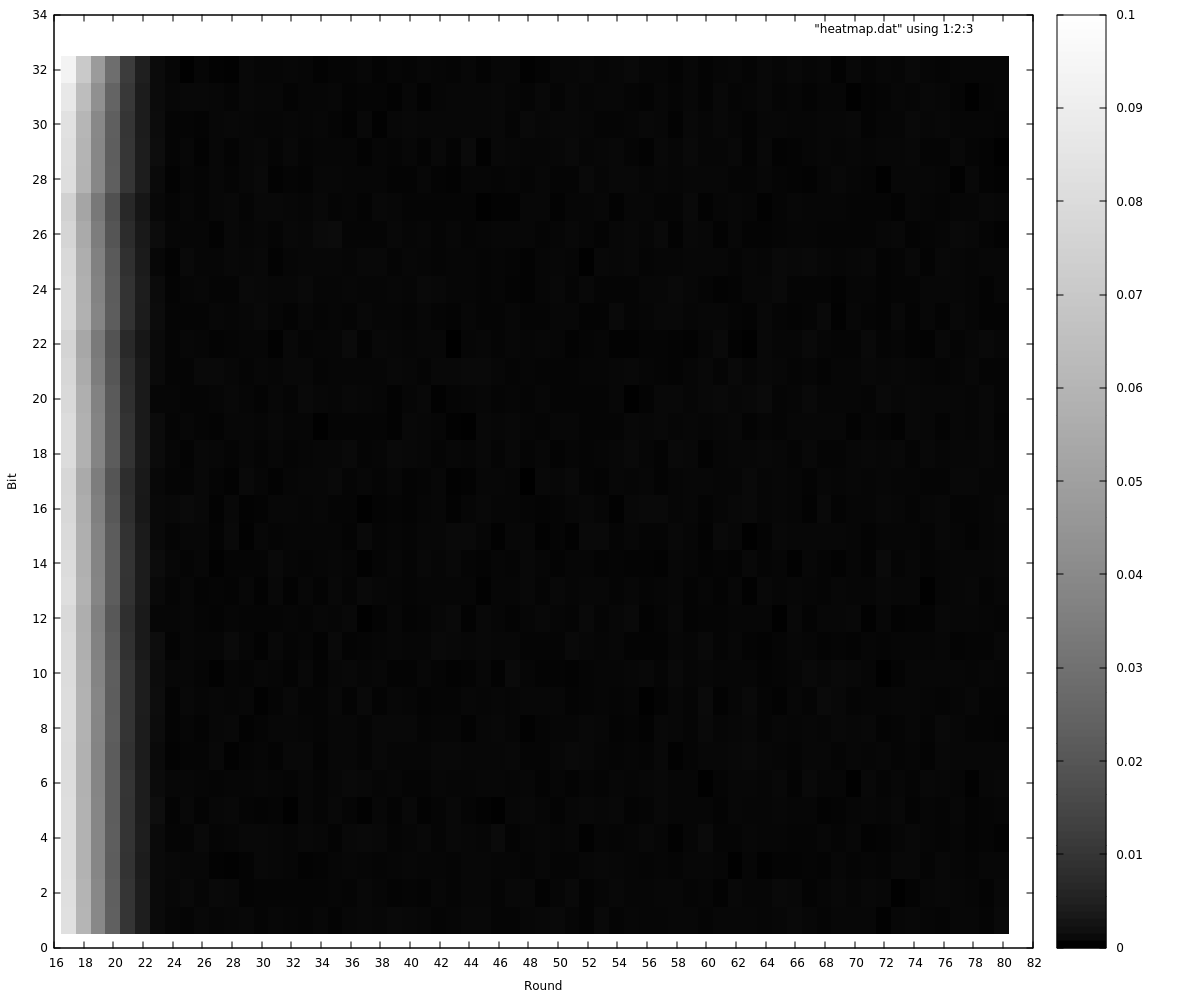}
	\caption{Divergence from SAC, round 17..80}
	\label{fig:divergence_17:80}
\end{figure}

As shown by Figure~\ref{fig:divergence_17:80}, the SHA-1 hash diverges from the SAC by remarkably small amounts.  The initial divergence is due entirely to the fact that the very last bits of a 672-bit input are found in rounds 15 and 16 and, when modified, have an exaggerated effect on subsequent rounds.  This effect is largely due to the fact that the changes have not yet had time to diffuse through the rounds.  Data which is most representative of the final hash output can therefore be seen in rounds $\geq 24$.

\begin{figure}[h]
	\centering
	\includegraphics[width=0.5\textwidth]{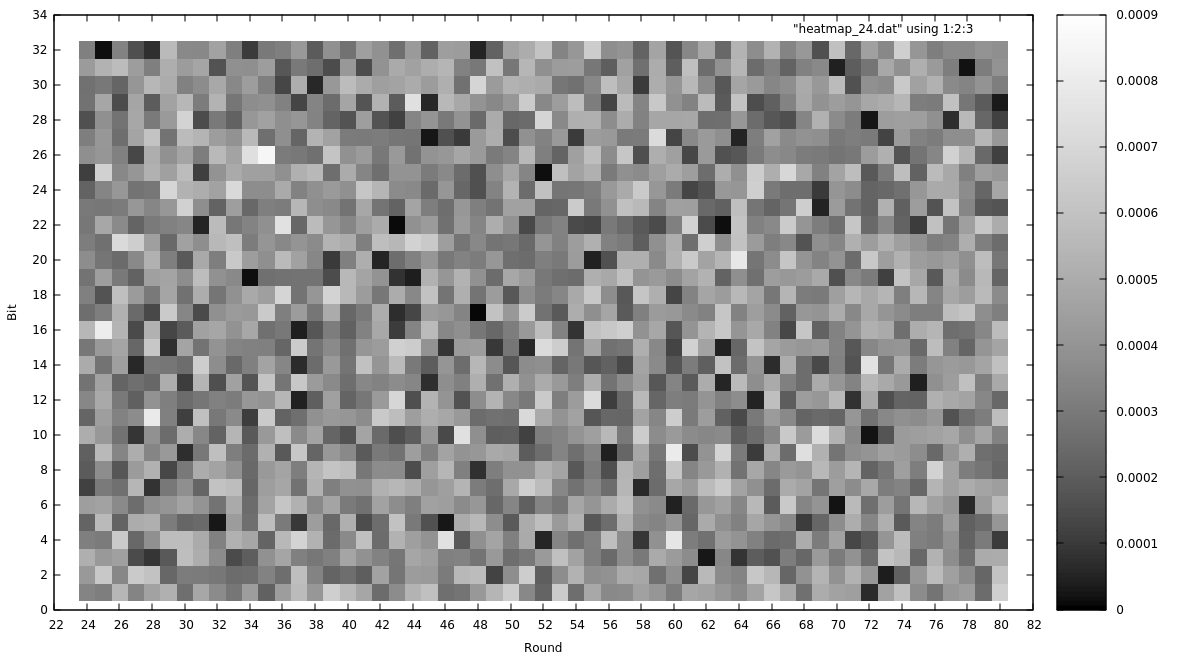}
	\caption{Divergence from SAC, rounds 24..80}
	\label{fig:divergence_24:80}
\end{figure}

If sufficient time is provided for diffusion, a different picture emerges.  Figure~\ref{fig:divergence_24:80} shows the absolute divergence from round 24 onwards.  Although the heatmap looks noisier, the most important thing to note is that the maximum divergence from the ``ideal'' SAC value of 0.5 is only 0.0009, which is within the margin of error for this sample size.

\begin{figure}[h]
	\centering
	\includegraphics[width=0.5\textwidth]{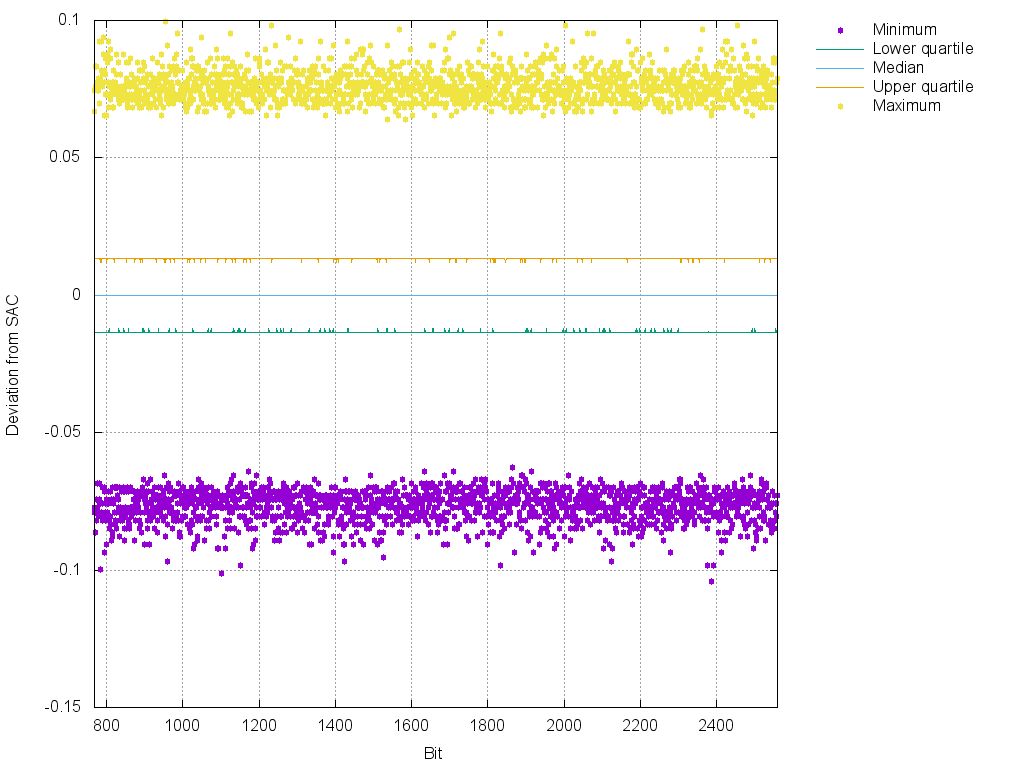}
	\caption{Summary statistics}
	\label{fig:summarystats}
\end{figure}

A 5-figure statistical summary (minimum, lower quantile, median, upper quantile, and maximum) of deviation from the SAC is plotted as Figure~\ref{fig:summarystats}.  In this graph, (round, bit) tuples have been converted to single value bits using the function $\mathit{bit}(r,i) = (r-1) \cdot 32 + (i-1)$.  This was done to better illustrate noteworthy points, and because there is no round-specific pattern in the data.  The median value is 0.0 throughout, and the lower and upper quartiles demonstrate remarkable consistency across the rounds despite minima and maxima which fluctuate significantly.  It is interesting to note that rounds 24..44 show the same pattern as rounds 60..80, which are the final rounds of the hash.  The distribution of values appears to remain constant from round 24 all the way up to round 80.

\begin{figure}[h]
	\centering
	\includegraphics[width=0.5\textwidth]{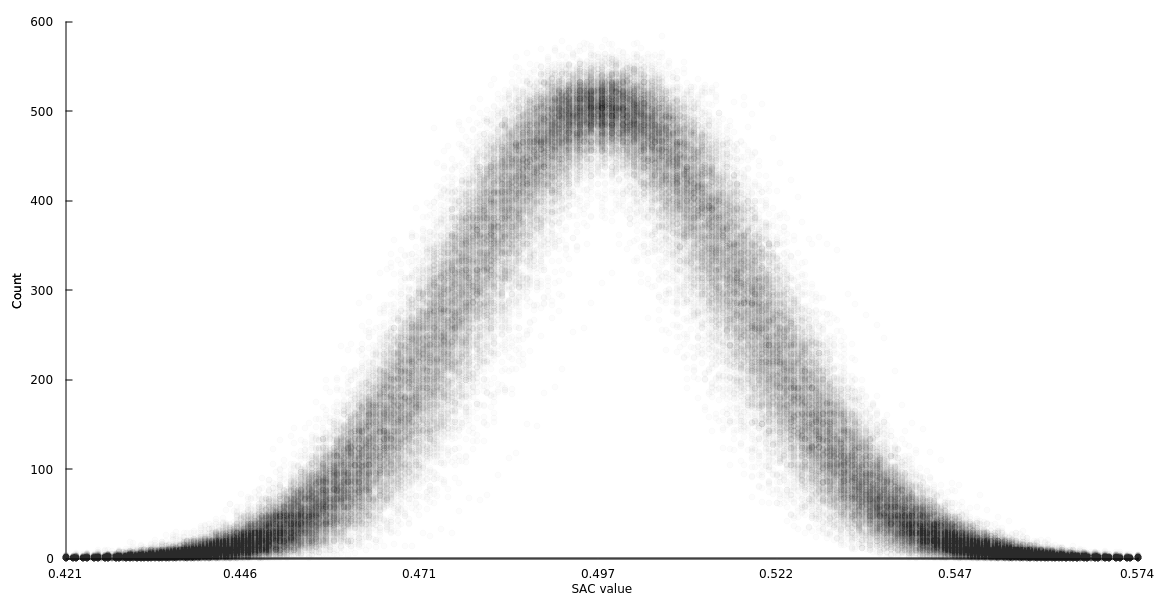}
	\caption{Distribution of SAC values}
	\label{fig:2dhisto}
\end{figure}

The distribution of SAC values for rounds $\geq 24$ is shown in Figure~\ref{fig:2dhisto}, and there are few surprises here.  It has a median, mean, and mode of 0.5, and appears to be a normal distribution.  To verify whether the distribution is, in fact, normal, a quantile-quantile plot was generated. A quantile-quantile plot overlays points from a data-set on top of the theoretically-predicted distribution; if the actual points lie along the theoretically-predicted line, then the data fits the specified distribution.

\begin{figure}[h]
	\centering
	\includegraphics[width=0.5\textwidth]{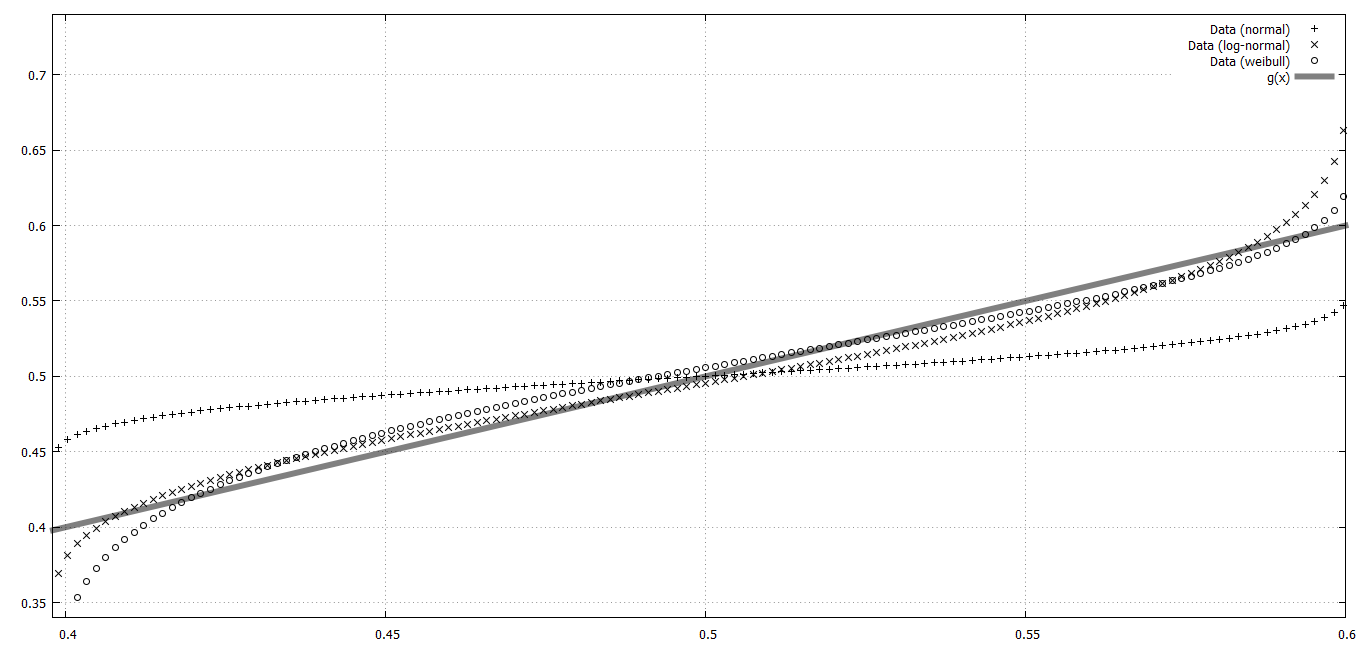}
	\caption{Quantile-quantile plot showing goodness of fit}
	\label{fig:qqplot}
\end{figure}

Three possible distributions were plotted (see Figure~\ref{fig:qqplot}):

\begin{itemize}
\item Normal ($\sigma=0.019285397$, $\mu=0.5$), using the standard deviation and mean of the data where round $\geq 24$.
\item Log-normal ($\sigma=0.059899039$, $\mu=0.49855239$), estimated from the data.
\item Weibull ($k=9.6116811$, $\lambda=0.52480750$), estimated from the data.
\end{itemize}

None of the distributions match the data exactly; in fact, the normal distribution is the worst fit, with log-normal and Weibull distributions being much closer fits.  At present, the distribution that the data conforms to is unknown.

\begin{figure}[h]
	\centering
	\includegraphics[width=0.5\textwidth]{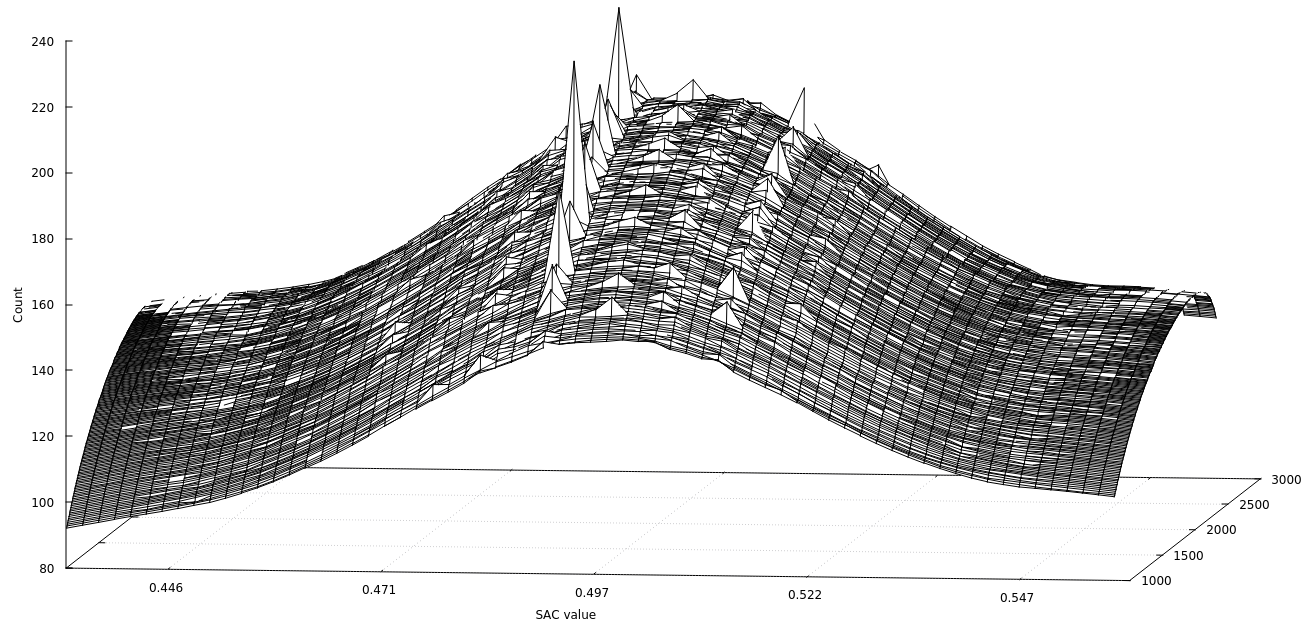}
	\caption{Distribution of SAC values, per-round}
	\label{fig:3dhisto}
\end{figure}

Lastly, the averaged per-round distribution of SAC values is shown as Figure~\ref{fig:3dhisto}, which may be regarded as a 3D histogram where zero-buckets have been discarded.  This graph attempts to show trends and changes in SAC values through rounds.  The ``spikiness'' of the center is immediately noticeable, despite the SAC values being averaged.  This reflects the fact that SAC values tend strongly towards 0.5.  The right side of the graph is shorter than the left, and also higher; since zero-buckets have been discarded, this would indicate that SAC values tend to be distributed into more buckets as they tend towards zero, and conversely concentrated into fewer buckets as they tend towards 1.  This tendency is present throughout the rounds.

The z-axis curve in the middle of the bits, which appears to be too pronounced to be an artifact of averaging, would seem to indicate that there are more non-zero buckets that are being filled as the bit-value increases.  However, such an increase should result in a decrease of SAC values at other points --- and a corresponding dip at those points on the x-axis.  There is no such dip, and other visualisations do not indicate such an increase.  In the absence of any other explanation, it is believed that it is an artifact of the visualisation.

\section{Discussion}

The questions posed may now be answered.  As the experimental results show, each of the output bits meets the SAC by round 24, and it therefore takes only 8 rounds from the end of the input data for the SAC values to settle into a ``stable'' state.  This stable state persists through all of the remaining rounds.

The distribution of SAC values does not fit any of the tested distributions exactly.  However, the distribution displays a regularity that makes it quite possible that a less well-known distribution will fit.  Further analysis of this could be worthwhile, since the distribution of SAC values may provide a different way to understand the behaviour of the hash.

One of the characteristics of a cryptographic hash is (second-)preimage resistance: the computational infeasibility of finding an input that results in a particular output.  The SAC results obtained from these experiments highlight the difficulty of obtaining a specific preimage since, from round 24 onwards, the SAC is either met or very closely approximated.  This makes it extraordinarily difficult to determine which input bit could contribute to a particular output change, since the answer is likely to be ``any of them''!

The methodology that has been described above is not specific to the SHA-1 hash, and may be applied to any hash function.  It would be interesting to see it applied to other hash functions with a view to comparing their SAC values and distributions to the results above.  Similarities and differences, and the possible reasons for them, would make for interesting research.  For example, SHA-1's spiritual predecessor, MD5, has also proven to be resistant to preimage attacks; could the reason be that it shares a similarly rapid achievement of close-to-SAC bits, followed by a similarly ``stable'' maintenance of the SAC through all of its rounds?

On an implementation note, it may be worthwhile to use a cloud computing platform (such as Google's BigTable) for future experiments of this nature.  The experiments have generated tens of gigabytes of data which take some time to query on a single machine.  The scalable infrastructure of the cloud may allow queries, and hence experiments, to proceed more quickly.




\bibliography{bibliography/converted_to_latex}

\end{document}